\documentstyle[12pt]{article}

\def\p{{\rm {\bf p}}}
\def\q{{\rm {\bf q}}}

\def\exp{{\rm exp}}
\def\g{\mbox{\boldmath$\gamma$}}

\begin{document}

\vspace{1.5cm}

\begin{center}

{\bf FERMI SPECTRA AND THEIR GAUGE INVARIANCE IN HOT AND DENSE ABELIAN
AND NON-ABELIAN THEORIES}

\vspace{1.5cm}

{\bf O.K.Kalashnikov}
\footnote{ Permanent address: I.~E.~Tamm Theoretical Physics
Division, P.~N.~Lebedev Physical Institute, Russian Academy of Science,
117924 Moscow, Russia; E-mail address:kalash@td.lpi.ac.ru}

High-Energy Physics, ICTP

34100 Trieste, Italy

\vspace{2.5cm}

{\bf Abstract}

\end{center}

The one-loop Fermi spectra (one-particle and collective ones) are found
for all momenta in the $T^2$-approximation and their gauge invariance
in hot and dense Abelian and non-Abelian theories is studied. It is
shown that the one-particle spectrum, if the calculation accuracy is
kept strictly, is gauge invariant for all momenta and has two branches
as the bare one. The collective spectrum always has four branches which
are gauge dependent including also their $|\q|=0$ limit. The exception
is the case $m,\mu=0$ for which this spectrum is gauge invariant for
all momenta as well.

\newpage

\section{Introduction}
The problem to obtain the gauge invariant results from gauge theories
(especially from non-Abelian ones) is not new but is very actual for
present-day physics. This problem is very many-sided and one easily
calls many tasks which require the first rate solution. In particular,
many questions arise when the effective electron (or quark) mass is
calculated within hot and dense QED or QCD, including even the pure QFT
(quantum field theory). Of course, the latter case is more studied but
some problems (in the first turn, the proof that the fermion mass is
infrared-finite) were solved only recently making calculation of the
perturbative mass in QFT selfconsistent and proving its gauge
invariance [1]. The completely different situation takes place when T
(temperature) is nonzero. In this case the nonzero thermal mass (or
thermal gap) is always generated [2,3]) and all Bose and Fermi spectra
(if we speak about so called collective ones) are split demonstrating
themselves more pronounced than in QFT [2-9]. The more complicated
scenario arises when the bare fermion mass $m$ and temperature are
nonzero simultaneously [10,11]. In this case the spectrum has the
additional splitting and changes its long wavelength asymptotical
behaviour: it becomes $\q^2$ instead of $|\q|$ when $m=0$. If along with
$m,T\ne 0$ the chemical potential $\mu$ is also nonzero the scenario
becomes very cumbrous [12,13] and only the separate spectrum limits can
be found analytically. To obtain the spectrum curves for all $|\q|$ the
numerical calculations are necessary that, unfortunately, hides many
details from the further analysis.  However, if scenarios with $\mu,T\ne
0$ and with $m,T\ne 0$ are considered separately the spectrum curves for
all $|\q|$ can be found analytically in any case within
$T^2$-approximation [14,15]. These analytical expressions easily
reproduce all known limits and are more convenient to investigate the
spectrum gauge invariance and many other properties. For the
general case when $\mu,m$ along with $T$ are nonzero only the effective
thermal mass can be found analytically [15,16] and being calculated
within different gauges at once demonstrates its gauge dependence.
However the situation drastically changes if the one-particle
spectrum [16] is considered instead of the collective one.  This
spectrum is analogous to the bare one but its properties are modified
due to interaction with medium and completely different from the
collective ones. Unlike the latter scenario the one-particle effective
mass is gauge invariant in the leading $e^2$-order (and, possibly, in
the higher orders as well) and has a real physical sense. The
nonperturbative fermion mass (the $|\q|=0$-limit of the collective
spectrum) is always gauge dependent as well as the full collective
spectrum, and the additional resummation seems to be necessary.

Briefly speaking the question of the gauge invariance for many results
found in statistical QCD is insufficiently studied and opens for
discussion. Even in statistical QED the exact gauge invariance is only
known for the photon thermal mass, but the same as in QCD, no exact
results exist for the thermal fermion masses and each times their gauge
invariance is necessary to investigate independently.

The goal of the present paper is to show that not only the perturbative
mass but also the full one-particle Fermi spectrum for all $\q^2$ is
gauge invariant if the calculation accuracy is kept strictly. This
invariance takes place for any $m,\mu$-parameters, although the case
$\mu=0$ is separated. In the case $\mu=0$ the full one-particle Fermi
spectrum is gauge invariant even beyond the perturbative accuracy in
any case within the simplest summation of one-loop calculations. In
the section 4 the collective spectrum is calculated for the case
$m,\mu\ne 0$ in the Coulomb gauge to compare it with one found in
the Feynman gauge.  It is shown that this spectrum is always gauge
dependent but, the Coulomb gauge seems to be more physical one for any
applications. The exception is the case $m,\mu=0$ [3,5] for which this
spectrum is gauge invariant for all momenta including their effective
mass. This scenario, probably, is valid in all perturbative orders
since in this case within $T^2$-approximation no dimensional
parameters (except $T$) are present and due to general theorems
[17,18], such spectrum should be gauge invariant exactly (including
all higher order corrections) even within hot QCD. Of course, the
Abelian QED and non-Abelian QCD should generate the different gauge
dependence of any spectra, but in the leading $e^2$ (or $g^2$) order
QED and QCD Fermi spectra are the same (if everywhere the numerical
factor is changed due to the prescription $e^2 \rightarrow
g^2(N^2-1)/2N$). Today there is a problem to find (or, at least,
understand) what a difference arises within the spectrum found in QED
when the higher order corrections are taken into account to distinguish
QED from QCD. The strong infrared divergencies of hot QCD (see, e.g.
[19] and other references within it) should display themselves via the
higher order corrections and it is not excluded that effective thermal
mass in QCD becomes the infrared-unstable. To start we choose hot and
dense QED (not QCD) although all results are the same in the
approximation considered. We also propose that everywhere the damping
is small and can be neglected considering that this question should be
investigated separately [20,21]. Our attention will be focused to solve
the fermion dispersion relation in a more complete form and to
investigate the gauge invariance of all spectra found.

\section{The one-loop fermion self-energy in the Coulomb gauge}
The Coulomb gauge is rather reliable gauge for perturbative
calculations within hot gauge Abelian and non-Abelian statistical
theories.  It does not generate (unlike the $\alpha$-gauges) the
additional infrared divergencies and requires only the standard
ultraviolet regularization.  Of course, all calculations in the Coulomb
gauge are more complicated than in the Feynman one but these are the
technical difficulties which can be overcome without any principle
modifications of the theory studied. In what follows our calculations
are performed in the Coulomb gauge but often we compare them with
the ones made in the Feynman gauge [14-16] and keep the same
abbreviations. Some details of these calculations and their cumbrous
results are placed in Appendix A.

Within the one-loop approximation the exact decomposition for
$\Sigma(q)$ is given by
\begin {eqnarray}
\Sigma(q)=i\gamma_{\mu}K_{\mu}(q)+m\;Z(q)
\end{eqnarray}
where three new scalar functions are introduced to find
nonperturbatively the function G(q)
\begin{eqnarray}
G(q)=\frac{-i\gamma_{\mu}({\hat q_{\mu}}+K_{\mu})+m\;(1+Z)}
{({\hat q_{\mu}}+K_\mu)^2\;+\;m^2\;(1+Z)^2} \,.
\end{eqnarray}
This representation leads to the one-loop dispersion relation (the
exact one has the additional functions [22]) for the Fermi excitations
which in any gauge has the form
\begin{eqnarray}
[\;(iq_4-\mu)-{\bar K}_4]^2\;=\;\q^2\;(1+K)^2 +m^2(1+Z)^2
\end{eqnarray}
and after the standard analytical continuation it can be solved
analytically or numerically. Here $K_4=i{\bar K_4}$ and
${\hat q}=\{(q_4+i\mu),\q\}$.

The calculations are rather lengthy and require at first to extract the
functions $Z(q)$ and $K_\mu(q)$ from $\Sigma(q)$
\begin{eqnarray}
\Sigma^C(q)&=&-e^2\int\frac{d^3p}{(2\pi)^3}\;\left\{\;\Bigr[
\frac{n_\p^+}{\epsilon_\p}\;
\frac{\gamma_4\epsilon_\p+i\g(\p-\q)\;[(\p-\q|\p)/(\p-\q)^2]+m}
{[q_4+ i(\mu+\epsilon_\p)\;]^2+(\p-\q)^2}\;\right.\nonumber\\
&+&\left.\;\frac{n_\p^B}{|\p|}\;
\frac{(|\p|+\mu-iq_4)\gamma_4-i\g\p\;[(\p|\p+\q)/\p^2]-m}
{[q_4+ i(\mu+|\p|)\;]^2+\epsilon_{\p+\q}^2}\;\right.\nonumber\\
&-&\left.\;\frac{n_\p^+}{\epsilon_\p}\;
\frac{\gamma_4\epsilon_\p-i\g\p-m}{2(\p-\q)^2}\;\Bigr]
-\Big[h.c.;(m,\mu)\rightarrow-(m,\mu)\Big]\right\}
\end{eqnarray}
and then to calculate them using a number of exact integrals. Our
result for the functions $Z(q)$ and $K_4(q)$ has the form
\begin{eqnarray}
&&Z^C(q_4,\q)=e^2\int\limits_{0}^{\infty}
\frac{d|\p|}{4\pi^2}\;\frac{|\p|}{2|\q|}\left\{\;\frac{1}{\epsilon_\p}
\Bigr[\frac{n_\p^++n_\p^-}{2}\;\Bigr(\;\ln(a_F^+a_F^-)\right.\nonumber\\
&&\left.-\ln\frac{(|\p|+|\q|)^2}{(|\p|-|\q|)^2}\;\Bigr)
+\frac{n_\p^+-n_\p^-}{2}\;\ln(\frac{a_F^+}{a_F^-})
\Bigr]-\frac{n_\p^B}{|\p|}\;\ln(a_B^+a_B^-)\right\}
\end{eqnarray}
\begin {eqnarray}
&&\!\!\!\!\!\!\!\!\!\!iK^C_4(q_4,\q)= e^2\int\limits_
{0}^{\infty}\frac{d|\p|}{4\pi^2}\frac{|\p|}{2|\q|}\left\{\;\Bigr[
\;\frac{n_\p^++n_\p^-}{2}\;\ln(\frac{a_F^+}{a_F^-})
+\frac{n_\p^+-n_\p^-}{2}\;\Bigr(\;\ln(a_F^+a_F^-)\right.\nonumber\\
&&\!\!\!\!\!\!\!\!\!\!\left.
+\ln\frac{(|\p|+|\q|)^2}{(|\p|-|\q|)^2}\;\Bigr)
\;\Bigr]+n_\p^B\;\Bigr[\;\ln(\frac{a_B^+}{a_B^-})+
\frac{(\mu-iq_4)}{|\p|}\;\ln(a_B^+a_B^-)\Bigr]\right\}
\end{eqnarray}
and the more complicated calculations are necessary to obtain
the vector $K_n(q)$ where $n=1,2,3$. Here we use a definition
$K_n(q)=q_n K(q)$ and after that the scalar function $K(q)$ is
calculated to be
\begin {eqnarray}
&&\!\!\!\!\!\!\!\q^2\;K^C(q_4,\q)=e^2
\int\limits_{0}^{\infty}\frac{d|\p|}{2\pi^2}\left\{\;\frac{\p^2}
{\epsilon_\p}\Biggr(\;\frac{n_\p^+-n_\p^-}{2}\;\Bigr[\;\frac{1}
{8|\p||\q|}\;\Bigr(h_F\ln(\frac{a_F^+}{a_F^-})\right.\nonumber\\
&&\!\!\!\!\!\!\!\!\left.+d_F\ln(a_F^+a_F^-)\Bigr)\Bigr]+\frac{n_\p^+
+n_\p^-}{2}\;\Bigr[\;1+\;\frac{1}{8|\p||\q|}\Bigr(\;h_F\ln(a_F^+a_F^-)
+d_F\ln(\frac{a_F^+}{a_F^-})\;\Bigr)\Bigr]\;\Biggr)\right.\nonumber\\
&&\!\!\!\!\!\!\!\!\left.+n_\p^B\;|\p|\;\bigr[1+\frac{1}{8|\p||\q|}
\Bigr(h_B\;\ln(a_B^+a_B^-)+d_B\;\ln(\frac{a_B^+}
{a_B^-})\;\Bigr)\;\Bigr]\right\}\;+\;\q^2{\tilde K}^C(q_4,\q)
\end{eqnarray}
where $h_F=\q^2-m^2-(iq_4-\mu)^2$ and $d_F=2\epsilon_\p(iq_4-\mu)$
(analogously $h_B=\q^2+m^2-(iq_4-\mu)^2$ and $d_B=2|\p|(iq_4-\mu)$)
are some simple abbreviations and other ones have the more complicated
form
\begin{eqnarray}
a_F^{\pm}&=&\frac{\q^2-m^2-(iq_4-\mu)^2 \pm
2\epsilon_\p(iq_4-\mu)-2|\p||\q|} {\q^2-m^2-(iq_4-\mu)^2 \pm
2\epsilon_\p(iq_4-\mu)+2|\p||\q|}
\end{eqnarray}
\begin{eqnarray}
a_B^{\pm}&=&\frac{\q^2+m^2-(iq_4-\mu)^2
\pm 2|\p|(iq_4-\mu)-2|\p||\q|} {\q^2+m^2-(iq_4-\mu)^2 \pm
2|\p|(iq_4-\mu)+2|\p||\q|}
\end{eqnarray}
Here and everywhere $\epsilon_\p=\sqrt{m^2+\p^2}$. The function
${\tilde K}^C(q_4,\q)$ has the rather cumbrous form and is placed
in Appendix A. This function is mainly essential beyond the
$T^2$-approximation and below only a few of its terms are
exploited to find $\mu/T$-corrections.

\section{The one-particle spectrum of massive Dirac particles in the
$T^2$-approximation}
The one-particle spectrum is the perturbative one and corresponds to the
calculations when in the leading order the dispersion relation (3) is
solved with $\Sigma(q_4,\q)$ taken at once on the bare mass shell
$iq_4=\mu \pm \sqrt{\q^2+m^2}$.

Within $e^2$-approximation this spectrum can be presented as follows:
\begin{eqnarray}
iq_4=(\mu+{\bar K}_4) \pm
\sqrt{m^2\Bigr(1+Z(\q)\Bigr)^2+\q^2\Bigr(1+K(\q)\Bigr)^2}
\end{eqnarray}
where all functions being put at once on the bare mass shell are
independent on $iq_4$. This spectrum is qualitatively different from
the collective one and at the beginning is more useful for many
applications. In particular, its long wavelength limit ($|\q|=0$-limit)
which defines the effective thermal mass is the gauge invariant value
[16]
\begin{eqnarray}
iq_4=\mu_R\;\pm\;m_R=\mu\;(\;1\;+\;2\; {\tilde I}_B\;)
\;\pm\;\Bigr[\;m(1-4I_Z)+\frac{I_K}{m}\;\Bigr]
\end{eqnarray}
At any rate Eq.(11) is the same in both gauges: the Coulomb and Feynman
ones. Here $I_B=\mu{\tilde I}_B$ and other abbreviations are:
\begin{eqnarray}
&&I_K=I_K^F+I_K^B=
\;e^2\int\limits_0^{\infty}\frac{d|\p|}{2\pi^2}
\;\epsilon_\p \;\frac{n_\p^++n_\p^-}{2}\;+\;
e^2\int\limits_0^{\infty}\frac{d|\p|}{2\pi^2}\;
|\p|\,n_\p^B \,,\nonumber \\ &&I_B=-e^2\int\limits_0^{\infty}
\frac{d|\p|}{4\pi^2}\;\frac{n_\p^+-n_\p^-}{2} \,,\qquad
I_Z=e^2\int\limits_0^{\infty}
\frac{d|\p|}{4\pi^2}\;\frac{n_\p^++n_\p^-}{2\epsilon_\p}\;.
\end{eqnarray}
The found gauge invariance is a very important property of this spectrum
and our task is to investigate whether this property keeps for all
momenta or only some separated spectrum limits demonstrate it within
the $T^2$-approximation. However, checking this property one should be
careful exploiting Eq.(10), since it is given beyond the accepted
one-loop accuracy. The more reliable expression has the form
\begin {eqnarray}
iq_4=(\mu\pm \epsilon_\q)\;+\;{\bar K}_4(\q)\;\pm\;
\Bigr(\;\frac{m^2 Z(\q)}{\epsilon_\q}\;
+\;\frac{\q^2K(\q)}{\epsilon_\q}\;\Bigr)
\end{eqnarray}
where the one-loop corrections to the bare spectrum are given
strictly within the $e^2$-accuracy. Below we return to
this question once more to demonstrate the important result: the
gauge invariance is a priori broken if the perturbative series are
considered beyond the accepted accuracy.

\subsection{The fermion one-particle spectrum with $m,\mu\ne 0$ in
Coulomb and Feynman gauges and its gauge invariance}
Here the one-particle Fermi spectrum will be found for all momenta in
the most general case $m,\mu\ne 0$ using the expression for $\Sigma(q)$
obtained in the Feynman (F.G) and Coulomb (C.G) gauges.  Namely these
gauges are preferable in statistics since only they are reliable without
any additional regularization.

Going to the $T^2$-approximation, we simplify all logarithms which
define the above found integrals, keeping only the leading $T^2$-term
and $\mu/T$-corrections. Effectively this operation leads to the
following ansatz
\begin{eqnarray}
\ln(\frac{a_F^+}{a_F^-})\simeq {\cal L}_{\pm}^F(\q)=\frac{1}{2}
\ln\Biggr[\frac{{\displaystyle 1+\frac{2\q^2}{m^2}(1 \mp
\frac{\epsilon_\q}{|\q|})}} {{\displaystyle 1+\frac{2\q^2}{m^2}(1
\pm \frac{\epsilon_\q}{|\q|})}}\Biggr]^2 \,,\;\;\;\;
\ln(\frac{a_B^+}{a_B^-})\simeq {\cal L}_{\pm}^B(\q)=\ln\Biggr[\frac
{{\displaystyle 1 \mp \frac{\epsilon_\q}{|\q|})}}{{\displaystyle 1
\pm \frac{\epsilon_\q}{|\q|})}}\Biggr]^2
\end{eqnarray}
and also $\ln(a_F^+a_F^-)\simeq -4|\q|/|\p|$ and
$\ln(a_B^+a_B^-)\simeq 0$.  In Eq. (14) and everywhere the bottom
signs of ${\cal L}_\pm^{F/B}$-quantities correspond to the signs in
the bare spectrum $iq_4=\mu \pm \sqrt{q^2+m^2}$ which is inserted into
all integrals before any expansion. The derived ansatz allows to
calculate the $T^2$-terms and $\mu/T$-corrections for all functions
involved in Eq.(10) explicitly and to compare the spectra found in
different gauges. The results of the calculations made in the Coulomb
gauge
\begin{eqnarray}
&&Z^C(\q)=-4I_Z-\frac{{\cal L}_{\pm}^F(\q)}{2|\q|}\;I_B \,,\qquad {\bar
K}_4^C(\q)=-\frac{{\cal L}_{\pm}^F(\q)}{4|\q|}\;I_K^F- \frac{{\cal
L}_{\pm}^B(\q)}{4|\q|}\;I_K^B\nonumber\\
&&\q^2K_{\pm}^C(\q)=I_K\mp \epsilon_\q K_4^C(\q)
\end{eqnarray}
and in the Feynman one
\begin{eqnarray}
&&Z^F(\q)=Z^C(\q)-\frac{{\cal L}_{\pm}^F(\q)}{2|\q|} I_B
\,,\qquad {\bar K}_4^F(\q)=K_4^C-2I_B\nonumber\\
&&\q^2K_{\pm}^F(\q)=2mI_B\;(\frac{m{\cal L}_{\pm}^F(\q)}{4|\q|}
\pm\frac{\epsilon_\q}{m})\;+\;\q^2K_{\pm}^C(\q)
\end{eqnarray}
are very similar but are not equal if $\mu\ne 0$.  This means that
$\Sigma(q_4,\q)$ and its structure functions are, as a rule, gauge
dependent but this is not the case for the one-particle spectrum since
within Eq.(13) the algebraic cancellations are possible to restore the
spectrum gauge invariance.

Indeed within the Feynman gauge the one-particle spectrum, before
algebraic cancellations, has the form
\begin {eqnarray}
iq_4&=&(\mu\pm \epsilon_\q)\;+\;({\bar K}_4^C(\q)-2I_B)\;\pm \;\left\{
\;\frac{m^2}{\epsilon_\q}\Bigr(\;Z^C(\q)-\frac{{\cal
L}_\pm^F(\q)}{2|\q|}I_B\;\Bigr)\right.\nonumber\\
&+&\left.\frac{1}{\epsilon_\q}\;\Bigr[\;2mI_B\Bigr(\;\frac{m{\cal
L}_\pm^F}{4|\q|} \pm \frac{\epsilon_\q}{m}\;\Bigr)\;+\;\q^2K_\pm^C(\q)
\;\Bigr]\;\right\}
\end{eqnarray}
but can be easily simplified as follows
\begin {eqnarray}
iq_4=(\mu\pm \epsilon_\q)\;+\;{\bar K}_4^C(\q)\;\pm\;
\Bigr(\;\frac{m^2 Z^C(\q)}{\epsilon_\q}\;
+\;\frac{\q^2K_\pm^C(\q)}{\epsilon_\q}\;\Bigr)
\end{eqnarray}
that is exactly the one-particle spectrum in the Coulomb gauge, and
consequently the gauge invariance is restored.

However the cancellations are not complete if the one-particle spectrum
is presented by Eq.(10). In this case the spectrum found in the Coulomb
gauge
\begin{eqnarray}
iq_4=(\mu+{\bar
K}_4^C(\q))\pm\sqrt{m^2(1+Z^C(\q))^2+\q^2\Bigr[1+\frac{1}{\q^2}
\Bigr(I_K(\q)\mp \epsilon_\q{\bar K}_4^C(\q)\Bigr)\Bigr]^2}
\end{eqnarray}
is essentially different from the one found in the Feynman gauge
\begin{eqnarray}
&&\!\!\!\!\!\!\!\!\!\!\!iq_4=(\mu-2I_B+{\bar K}_4^C(\q))\pm\\
&&\nonumber\\
&&\!\!\!\!\!\!\!\!\!\!\!\!\sqrt{m^2(1+Z^C(\q)-\frac{{\cal
L}_{\pm}^F(\q)}{2|\q|}I_B)^2+\q^2\Bigr[1+\frac{1}{\q^2}
\Bigr(I_K(\q)\mp \epsilon_\q{\bar K}_4^C(\q)+
2mI_B(\frac{m{\cal L}_{\pm}^F(\q)}{4|\q|}
\pm\frac{\epsilon_\q}{m})\Bigr)\Bigr]^2}\nonumber
\end{eqnarray}
since the accepted accuracy is exceeded within this scenario.
These spectra are gauge dependent although their low and high energy
limits reproduce the gauge invariant results as well. The correct
one-particle spectrum should be found using Eq.(13) and has the form
\begin {eqnarray}
iq_4=(\mu\pm \epsilon_\q)\;
\pm  \Bigr[\;-\;\frac{m^2}{\epsilon_\q}\;\Bigr(\;4I_Z
+\;\frac{{\cal L}_\pm^F(\q)}{2|\q|}I_B\;\Bigr)\;+\;\frac{I_K}
{\epsilon_\q}\;\Bigr]
\end{eqnarray}
which is our main result for this section. To find its limit for
small $\q^2$ one should use the standard expansion of ${\cal
L}_{\pm}^{F/B}(\q)$-quantities
\begin{eqnarray}
{\cal L}_{\pm}^{F/B}(\q)=\mp \frac{4|\q|}{m}
\pm \frac{2|\q|^3}{3m^3}+O(\q^5)
\end{eqnarray}
that is the same for both functions in the leading order and perform
the simple algebra. These calculations yield the following result:
\begin{eqnarray}
iq_4\;=\;\mu\;(1+2{\tilde I}_B)\;\pm \;m_F\;
\pm \;[\;(\;1+4I_Z-\frac{I_K}{m^2}\;)
\mp \mu\;\frac{8{\tilde I}_B}{3m}\;]\;\frac{\q^2}{2m}\;+\;O(\q^3)
\end{eqnarray}
where $m_F=m(1-4I_Z)+I_K/m$ is the effective fermion mass accepted for
$e^2$-approximation.

The high energy limit in the $T^2$-approximation is also easily
calculated and has the form
\begin{eqnarray}
iq_4\;=\;\mu\;\pm \;|\q|\;\pm \;\frac{1}{|\q|}\;[\;I_K-4m^2I_Z\;]
\;+\;O(\frac{1}{\q^2}\ln\frac{\q^2}{2m^2})
\end{eqnarray}
which is analogous to the bare one. All other terms are small
and can be neglected. Comparing the low and high energy limits
we see that one spectrum branch (the case of an upper sign ) can
demonstrate the minimum at finite momentum if the density is rather
large.

\subsection{The gauge invariant one-particle spectrum with $\mu=0$ in
the $T^2$-approximation}
The special case of zero density (when $\mu=0$) is rather
interesting  since in this case within the $T^2$- approximation
there is no problem with the gauge invariance at all.  In this case the
one-particle Fermi spectrum is gauge invariant at the beginning and
Eq.(10) can be used to present it as follows
\begin{eqnarray}
iq_4\;=\;{\bar K}_4^{\pm}(\q)\;\pm
\;\sqrt{\;m^2(1-4I_Z)^2\;+\;\q^2[\;1+\frac{1}{\q^2}\;(\;I_K\mp
\epsilon_\q {\bar K}_4^{\pm}(\q)\;)\;]^2}
\end{eqnarray}
where
\begin{eqnarray}
{\bar K}_4^{\pm}(\q)=-\frac{{\cal L}_{\pm}^F(\q)}{4|\q|}\;I_K^F
-\frac{{\cal L}_{\pm}^B(\q)}{4|\q|}\;I_K^B \;=\;\pm \frac{I_K}{m}\;
\mp \frac{\q^2}{6m^3}\;I_K\;+\;O(\q^4)
\end{eqnarray}
Its low and high energy limits repeat the appropriate expressions found
above in the Coulomb gauge for the case $\mu=0$ but, and it is more
important, this spectrum is gauge invariant at once and, probably,
this property is kept within all higher order corrections

\section{The collective Fermi spectrum
and its gauge dependence in the $T^2$-approximation}
For hot gauge theory (Abelian or non-Abelian) with massless fermions
and in the symmetrical case ($\mu=0$) the Fermi excitations have
collective nature and appear as "massive" quasi-particles (or
quasi-holes) under the physical vacuum. In the simplest case their
thermal masses (or more exactly their thermal gaps) are the same for
each spectrum branch and they arise dynamically in spite of the exact
chiral symmetry inherent to initial Lagrangian on the operator level.
Such excitations in hot QCD were first found in [3] and then studied in
many other papers. Their spectrum [3,5] was found analytically in the
$T^2$-approximation for all momenta
\begin {eqnarray}
\omega_\pm^2(\xi)=\xi^2\;\omega_0^2\;\Bigr(\;
\frac{\eta} {\xi-\eta}+\frac{\eta}{2}\ln\frac{\xi-1}{\xi+1}\;\Bigr)
\,,\qquad \eta=\pm 1
\end{eqnarray}
and is gauge invariant in any case within one-loop
calculations.  Here the variable $\xi$ runs over the range
$1<\xi<\infty$ and the long wavelength spectrum limit (i.e.
$\q=0$-limit) corresponds to the case $\xi\rightarrow \infty$. In QCD
the thermal gap is: $\omega_0^2=g^2T^2/6$, and the spectrum is split
since $\eta=\pm 1$. All its branches have the finite gap at zero
momentum (the finite thermal mass) and $\omega_-^2(\xi)$ quasi-hole
branches have a very specific minimum at finite $\q$.  This minimum
always exists when $m,\mu=0$ and is a very interesting subject for
discussion [3-9].  For small $\q$ this spectrum has the linear limit
\begin {eqnarray}
\omega_\pm^2(\q)\;=\;\omega_0^2\;\Bigr[\;1\pm
\frac{2}{3}\;\frac{|\q|}{\omega_0}\;+\;\frac{7}{9}\;\frac{\q^2}
{\omega_0^2}\;+\;O(|\q|^3)\;\Bigr]
\end{eqnarray}
since all branches have the finite gap, but no real mass. In
the high momentum region the quasi-particle and quasi-hole branches
have qualitatively different limits. The quasi-particle branches
demonstrate the standard powerful behaviour
\begin {eqnarray}
\omega_+^2(\q)\;=\;\q^2\;+\;2\;\omega_0^2\;-\;(\frac{\omega^4}{\q^2})\;
\ln\frac{\q^2}{\omega_0^2}\;+\;O(\frac{1}{\q^2})
\end{eqnarray}
but this is not the case for the quasi-hole ones
\begin {eqnarray}
\omega_-^2(\q)\;=\;\q^2\;\Bigr[\;1+\;4\;
\exp\Bigr(-\frac{2\q^2}{\omega_0^2}\;\Bigr)\;+\;O\Bigr(
\exp\Bigr(-\frac{4\q^2}{\omega_0^2}\Bigr)\;\Bigr]
\end{eqnarray}
which more quickly (exponentially) approach the line $\omega=\q$.

However this scenario changes drastically when $m$ or $\mu$ is not
equal to zero. In the first turn this spectrum becomes gauge dependent
and more complicated: its mass (or gap) is split and its minimum
inherent to the quasi-hole branches, as a rule, disappears. We study
this situation below using the Coulomb gauge and compare our
calculations with the results found in the Feynman one from the papers
[14,16] where these excitations were also obtained analytically for all
momenta with nonzero $m,\mu$-parameters.

\subsection{The collective spectrum of massless Dirac particles
in the Coulomb gauge for the case $\mu\ne 0$}
The scenario with $m=0$ but $\mu\ne 0$ is very similar to the previous
one where $m,\mu=0$ and keeps many its properties although the symmetry
$(iq_4 \rightarrow -iq_4)$ is lost. Here, as previously, all
spectrum branches have the finite gap and the linear limit for small
$\q$, but there are two different gaps one of which, sometimes, can be
equal to zero.

The dispersion relation (3) is simplified to be
\begin {eqnarray}
[\;(iq_4-\mu)-{\bar K}_4]^2\;=\;\q^2\;(1+K)^2
\end{eqnarray}
and its solution can be written as follows
\begin {eqnarray}
(iq_4-\mu)-{\bar K}_4\;=\;\eta |\q|\;(1+K)
\end{eqnarray}
Here $\eta=\pm 1$ and we use the simple redefinition $K_4=i{\bar
K}_4$.  All functions involved into Eq.(32) are calculated using
the $T^2$-approximation where the essential simplifications are
possible due to the simple ansatz
\begin {eqnarray}
\ln(a^+a^-)=-\frac{2|\q|}{|\p|}+O(\frac{1}{T^2})\,,\qquad
\ln\frac{a^+}{a^-}=2\ln\frac{\xi-1}{\xi+1} + O(\frac{1}{T^2})
\end{eqnarray}
which keeps only $T^2$-terms and $\mu/T$-corrections in Eq.(32). With
the ansatz (33) the further calculations are easily performed and their
results are given by

\begin {eqnarray}
&&{\bar K}^C(q_4,\q)\;=\;\frac{I_K}{\q^2} \Bigr(\;1+\frac{\xi}{2}\ln
\frac{\xi-1}{\xi+1}\Bigr)\;-\;I_B\;\Bigr(\;\xi-\frac{1}{2}
(1-\xi^2)\ln\frac{\xi-1}{\xi+1}\;\Bigr)\;\frac{1}{|\q|}\nonumber\\
&&{\bar K}_4^C(q_4,\q)=I_B-\frac{I_K}{2|\q|}
\ln\frac{\xi-1}{\xi+1}
\end{eqnarray}
where all numerical integrals $I_K$ and $I_B$ are defined by Eq.(12).
Now one should plug the expressions found above into Eq.(32) and
perform a simple algebra to find $\omega=\xi|\q|$. Here
$\omega=(iq_4-\mu)$ and the variable $\xi$ is more convenient than
$|\q|$. The latter should be excluded from Eq.(32).  The result is the
quadratic equation with respect to $\omega(\xi)$
\begin {eqnarray}
\omega^2\;(\;\xi-\eta\;)\;-\;\omega\;\xi\;I_B\;(\;1-\eta\;b(\xi)\;)
\;-\;I_K\;\xi^2\;A(\xi)=0
\end{eqnarray}
where the functions $A(\xi)$ and $b(\xi)$ are given by
\begin {eqnarray}
A(\xi)=\eta \;(1+\frac{\xi-\eta}{2}\ln\frac{\xi-1}{\xi+1}\;)\nonumber\\
b(\xi)=\xi-\frac{1}{2}(1-\xi^2)\ln\frac{\xi-1}{\xi+1}
\end{eqnarray}
Keeping the $e^2$-accuracy our solution of Eq.(35) is found to be
\begin {eqnarray}
\omega_\pm(\xi|\eta)\;=\;\xi\;\Bigr[\;\frac{I_B\;(1-\eta\;b(\xi)\;)}
{2(\xi-\eta)}\;\pm \sqrt{I_K\;\Bigr(\;\frac{\eta}
{\xi-\eta}+\frac{\eta}{2}\ln\frac{\xi-1}{\xi+1}\;\Bigr)}\;\;\;\Bigr]
\end{eqnarray}
and presents the one-loop spectrum in medium for the case $\mu\ne 0$.
Here $\eta=\pm 1$ as usual and $\omega=iq_4-\mu$. The spectrum
(37) has four branches which are split for all momenta, excepting
their $|\q|=0$ limit, where they are combined by pairs demonstrating
two different gaps
\begin {eqnarray}
E_\pm=\frac{I_B}{2}\;\pm \sqrt{I_K}
\end{eqnarray}
Their asymptotic behaviour for small $|\q|$ has the form
\begin {eqnarray}
\omega_\pm(\q|\eta)\;=\;E_\pm+\frac{\eta}{3}\;|\q|
+\frac{1}{6}\;\Bigr(\;I_B\;\pm \;\frac{8}{3}\;\sqrt{I_K}\;\Bigr)\;
\frac{\q^2}{E_\pm^2}\;+\;O(\q^3)
\end{eqnarray}
where one branch always has the minimum at finite momentum.
Two branches, if $E_-=0$, lose their gap, but this can occur only
under the special constrain on $\mu,T$-parameters.

Unfortunately this scenario is gauge dependent and spectrum (37)
differs from the one found in the Feynman gauge [14]. Nevertheless,
some mathematical correspondence takes place: the spectrum in the
Feynman gauge arises from Eq.(37) if $\omega(\xi)$ will be replaced
to $[-\omega(\xi)]$. This correspondence is rather strange and is
inherent only to this simple scenario where the mixing is absent
between $\mu$ and $m$ parameters. If $\mu,m\ne 0$ the situation is
more complicated and the gauge invariance in this case seems
to be very problematical.

\subsection{The collective spectrum of massive Dirac particles
for the case $\mu=0$ in the Coulomb gauge}
It is another scenario which can be considered analytically for all
$|\q|$ in the $T^2$-approximation.  Here $I_B=0$ identically and
symmetry ($iq_4 \rightarrow -iq_4$) is restored. But now the
dispersion relation has the additional mass term
\begin {eqnarray}
[\;(iq_4-\mu)-{\bar K}_4]^2\;=\;\q^2\;(1+K)^2 +m^2(1+Z)^2
\end{eqnarray}
and all calculations are more complicated. To solve Eq.(40),
taking into account only the leading $m/T$-corrections, we use the
functions from Eq.(34) with $\mu=0$
\begin {eqnarray}
{\bar K}^C(q_4,\q)\;=\;\frac{I_K}{\q^2}
\Bigr(\;1+\frac{\xi}{2}\ln\frac{\xi-1}{\xi+1}\;\Bigr)\,,\qquad
{\bar K}_4^C(q_4,\q)=-\frac{I_K}{2|\q|}\ln\frac{\xi-1}{\xi+1}
\end{eqnarray}
and $Z^C(q_4,\q)=-3I_Z$. With these functions Eq.(40)
is transformed to be
\begin {eqnarray}
\omega^4\;(\;\xi^2-1\;)\;-\;\omega^2\;\xi^2\;[\;m^2(1-3I_Z)^2+2I_K\;]
\;-\;I_K^2\;\xi^4\;(\;1+\xi\ln\frac{\xi-1}{\xi+1}\;)\;=\;0
\end{eqnarray}
and becomes the quadratic equation in respect to $\omega^2(\xi)$ (not
to $\omega(\xi)$ as previously). However, this is not the case when
$m,\mu\ne 0$ simultaneously (see [15] and Appendix B). In this case the
dispersion relation generates the full equation of the fourth degree in
respect to $\omega(\xi)$ and requires the numerical calculations. The
analytical solution found within Eq.(42) has the form
\begin{eqnarray}
\omega_{\pm}(\xi)^2=\frac{\xi^2(\;2I_K+m_R^2)}{2(\xi^2-1)}\;
\pm\;\sqrt{\frac{\xi^4}{(\xi^2-1)^2}\Bigr[\;(b(\xi)I_K)^2
+\;m_R^2(\;I_K+m_R^2/4)\Bigr]}\;.
\end{eqnarray}
and presents the collective Fermi spectrum for all $|\q|$ in the
Coulomb gauge. Here $m_R^2=m^2(1-3I_Z)^2$ and, namely, this quantity
differs the Coulomb and Feynman gauges, where $m_R^2=m^2(1-2I_Z)^2$.
Excepting this quantity the found spectra are completely the same in
the both gauges, and, in some sense, are gauge invariant. It is not
excluded that the additional resummation is necessary to improve the
situation, but today this ansatz is unknown. The found spectrum is
rather complicated and presents four completely separated branches:
two branches (when the plus sign is taken in Eq.(43)) correspond to
quasi-particle excitations and the other two (when the minus sign is
taken) to quasi-hole ones.  These spectrum branches differ in
their asymptotic behavior and in many other properties.

The long-wavelength behavior of these spectrum branches (when
$\xi\rightarrow\infty$) has the form
\begin {eqnarray}
\omega_{\pm}(|\q|)^2=\;M_{\pm}^2\;+\;\Bigr(\;M_{\pm}^2\pm
\frac{4}{9}\frac{I_K^2}{\sqrt{\;m_R^2(m_R^2+4I_K\;)}}
\;\Bigr)\;\frac{|\q|^2}{M_{\pm}^2}\;+\;O(|\q|^4)
\end{eqnarray}
where the squares of the effective masses are given by
\begin {eqnarray}
M_{\pm}^2=\;\frac{m_R^2}{2}+I_K\pm
\sqrt{\;m_R^2\Bigr(\frac{m_R^2}{4}+I_K\;\Bigr)}\;.
\end{eqnarray}
These masses are different for all branches
$M_{\pm}=\frac{1}{2}(\eta m_R\pm \sqrt{m_R^2+4I_K})$ and are in
agreement with the results found in many other papers. Here $\eta=\pm
1$. It is also important that the quasi-hole branches $\omega_-(|\q|)^2$
are very sensitive to the choice of the parameters $m,T$.  In many
cases these branches are monotonic functions for small $|\q|^2$, and
the well-known minimum [3] disappears. Although this minimum always
exists for massless particles, the special conditions are necessary to
generate it when $m \ne 0$.

In the high-momentum region the asymptotical behaviors found for the
quasi-particles and quasi-hole excitations are also completely
different. The quasi-particle branches are approximated within a
rather usual expression
\begin{eqnarray}
\omega_{+}(|\q|)^2=\;|\q|^2\;+\;(2I_K+m_R^2)\;
-\;\frac{I_K^2}{|\q|^2}\ln\frac{4|\q|^2}{2I_K+m_R^2}\;
+\;O(\frac{1}{\q^2})
\end{eqnarray}
where the nonanalytic term is not essential. The situation is different
for the quasi-hole excitations, which do not exist in the vacuum
(when $T$ and $\mu$ are equal to zero). They disappear very rapidly, and
their asymptotic behaviour is found to be
\begin{eqnarray}
\omega_{-}(|\q|)^2=
\;|\q|^2\;\Bigr[\;1+4\;\exp(-\frac{|\q|^2(2I_K+m_R^2)}{I_K^2})
\;+\;O(\;\exp(-\frac{2|\q|^2(2I_K+m_R^2)}{I_K^2})\;\bigr]
\end{eqnarray}
In the high momentum region these spectrum branches approach the line
$\omega^2=|\q|^2$ more quickly than (46).

\subsection{The collective thermal mass in the
Coulomb and Feynman gauges and its gauge dependence}
In the most general case when $\mu,T$ and $m$ are nonzero
simultaneously, the effective mass can be also calculated analytically
[14-16]. Here we compare it with the results found above and discuss
its gauge dependence. In the Coulomb gauge this mass (see [16] and
Appendix B) is given by
\begin{eqnarray}
\omega_\pm^C(0)=\frac{1}{2}\Big[\eta\;m_R^C+I_B\Big]\pm
\sqrt{\;\frac{[\eta\;m_R^C+I_B]^2}{4} +(I_K+2\eta m I_B)} \,.
\end{eqnarray}
where $m_R^C=m(1-3I_Z)$. The mass found is split and has the
well-separated mass spectrum, where two values of them pertain to
the quasi-particle excitations and other two present the quasi-hole
ones.  This is evident from Eq.(48) where $\eta=\pm 1$.  However, this
spectrum is not the same as in the Feynman gauge [14]
\begin{eqnarray}
\omega_\pm^F(0)=\frac{1}{2}\Big[\eta\;m_R^F-I_B\Big]\pm
\sqrt{\;\frac{[\eta\;m_R^F-I_B]^2}{4} +(I_K+4\eta m I_B)} \,.
\end{eqnarray}
where $m_R^F=m(1-2I_Z)$ and the $\mu$-dependence is different. Thus,
these masses, excepting the case $m,\mu=0$, are gauge dependent and
this is true although Eqs.(48),(49) are given beyond the accepted
calculation accuracy. If one strictly keeps the $e^2$-accuracy these
equations are:
\begin{eqnarray}
\omega_\pm^C=\frac{1}{2}(\eta\pm 1)m_R^C\pm\frac{I_K}{m}\pm
\frac{I_B}{2}(5\eta\pm 1)\,,\qquad
\omega_\pm^F=\frac{1}{2}(\eta\pm 1)m_R^F\pm\frac{I_K}{m}\pm
\frac{I_B}{2}(7\eta\mp 1)
\end{eqnarray}
where we propose that $m\ne 0$. If at once $m=0$ one has the simpler
result
\begin{eqnarray}
\omega_\pm^C=\frac{I_B}{2}\;\pm \;\sqrt{I_K} \,,\qquad
\omega_\pm^F=-\frac{I_B}{2}\;\pm \;\sqrt{I_K}
\end{eqnarray}
which repeats Eq.(38). Eq.(50) leads to the following mass spectrum
\begin{eqnarray}
m_p^{(C/F)}\;=\;\eta\;(\;m_R^{(C/F)}\;+\;\frac{I_B}{m}\;)\;+\;3I_B
\end{eqnarray}
\begin{eqnarray}
m_h^C\;=\;\eta\;\frac{I_K}{m}\;-\;2I_B\,,\qquad
m_h^F\;=\;\eta\;\frac{I_K}{m}\;-\;4I_B
\end{eqnarray}
which demonstrates that the gauge invariance for the quasi-hole thermal
masses in (53) is destroyed more sharply than for the quasi-particle
masses in (52). It is very likely, that in this situation the futher
resummation easily restores the gauge invariance of the quasi-particle
thermal mass even if $\mu \ne 0$ but the problem with the quasi-holes
one, probably, remains.

\section{Conclusion} To summarize we have established two kinds of
fermion spectra in statistical QED and the same in QCD: the
one-particle spectrum and the collective one. The one-particle spectrum
is qualitatively the same as the bare one and is gauge invariant, in
any case within the one-loop approximation. It has two branches
$iq_4=\mu_R \pm m_R$ and their chemical potential and effective mass
are only quantitatively changed due to the interaction with the medium.
It is not excluded that within QED this spectrum is gauge invariant in
all higher orders as well, if the calculation accuracy is kept
strictly.  Moreover in hot QED this spectrum seems to be gauge
invariant exactly but this is not the case, a priori, for hot QCD where
the strong infrared divergencies spoil this scenario.

The collective fermion spectrum was calculated in the Coulomb gauge
and was compared with the one found in the Feynman gauge to investigate
its gauge dependence. This spectrum is additionally split (usually for
all momenta) and its branches always have the nozero thermal masses
different for all the spectrum ones, if $m\ne 0$. These masses arise
dynamically and are nozero even if the exact chiral symmetry forbids
them on the operator level.  They are not correlated with bare mass and
are generated always: but only for the case $m,\mu=0$ this thermal mass
and the full collective spectrum are gauge invariant. For any other
cases the collective spectrum is gauge dependent and its connection
with the real excitations should be investigated separately.

However one can see (comparing our results in the Feynman and Coulomb
gauges) that gauge invariance is broken variously in different
scenarios. This fact demonstrates itself more sharply when $\mu\ne 0$
[due to asymmetry ($iq_4 \rightarrow -iq_4$)] and is smoothly expressed
(only slightly shifts all spectrum branches) for the symmetric case
where $\mu=0$ and $\omega^2$ (not $\omega$) presents all the spectrum
ones. In the last case (namely, when $\mu=0$), the additional
resummation, probably, can change the situation to restore the gauge
invariance. But this is very unlikely for the scenario with $\mu\ne 0$
since in this case the spectra found (especially the quasi-hole ones)
are rather different and their rearrangement seems to be difficult.
Moreover this problem is more aggravated if all calculations are
performed beyond the $T^2$-approximations. In this case one at once
encounters many difficulties arising due to the ill-infrared
behaviour inherent to hot QCD (or QED as well) and all calculations
become very ambiguous. The coexistence of the infrared peculiarities of
hot gauge theories (like QCD) with the gauge invariance of their
results is very problematical and this remains a very serious problem
for any calculations performed today (especially for nonperturbative
ones).

The additional problems arise if the damping is taken into account
[20,21], especially when these calculations are performed beyond the
$T^2$-approximation. Here the selfconsistent calculations are
only acceptable, where, in the first turn, the calculation accuracy is
strictly kept, and the resummation scheme is checked to be, at least,
gauge covariant (the Ward identities should be satisfied between the
effective propagators and vertices).

\begin{center}
{\bf ACKNOWLEDGMENTS}
\end{center}

I am very grateful to S. Randjbar-Daemi for inviting me to the
International Center for Theoretical Physics in Trieste, and I am very
obliged to the entire staff of this center for their kind hospitality.

\newpage

\begin{center}
{\bf APPENDIX A}
\end{center}
The one-loop  fermion self-energy in the Coulomb gauge, after the
standard summation over the spinor indices and frequencies is presented
by Eq.(4). Its calculation strongly differs from the calculations
performed in the Feynman gauge only on the last stage when integration
over angles is performed.  This difference demonstrates itself more
pronounced only for the vector $K_n(q_4,\q)$ which has the form
\begin{eqnarray}
&&\!\!\!\!\!\!K_n^C(q_4,\q)=e^2\int\frac{d^3p}{(2\pi)^3}\;\left\{\;-\;
\frac{\p_n\;(n_\p^++n_\p^-)}{2\epsilon_\p\;(\p-\q)^2}\right.\\&&\!\!\!\!
\!\!\left.+\;\p_n\;\Big(\;1+\frac{\p\q}{\p^2}\;\Big)\;\frac{n_\p^B}{|\p|}
\;\Biggr(\;\frac{1}{(\;i\p+{\hat q}_4\;)^2+\epsilon_{\p+\q}^2}\;+\;
\frac{1}{(\;-i\p+{\hat q}_4\;)^2+\epsilon_{\p+\q}^2}\;\Biggr)\right.
\nonumber\\ &&\!\!\!\!\!\!\left.-\;\Big(\;1+\frac{\p-\q|\q}{(\p-\q)^2}
\Big)\;\frac{(\p-\q)_n}{\epsilon_\p}\;\Biggr(\;
\frac{n_\p^+}{(\;i\epsilon_\p+{\hat q}_4\;)^2
+(\p-\q)^2}\;+\;\frac{n_\p^-}{(\;-i\epsilon_\p+{\hat q}_4\;)^2
+(\p-\q)^2}\;\Biggr)\;\right\}\nonumber
\end{eqnarray}
and we propose that $K_n^C(q_4,\q)=q_nK^C(\q_4,\q)$. Then the very
lengthy but standard calculations lead to the result presented in
Eq.(7) where the function ${\tilde K}^C(q_4,\q)$ is given by
\begin{eqnarray}
\q^2\;{\tilde K}^C(q_4,\q)=I_1+I_2+I_3 \;,
\end{eqnarray}
\begin{eqnarray}
&&\!\!\!\!\!\!\!I_1=-e^2\int\limits_{0}^{\infty}
\frac{d|\p|}{4\pi^2}\;\frac{\p^2}{\epsilon_\p}\;
\frac{n_\p^+-n_\p^-}{2}\frac{1}{4|\p||\q|(\epsilon_\p^2
+{\hat q}_4^2)^2}\nonumber\\&&\left\{-(\p^2-\q^2)^2\;d_F\;
\ln\frac{(\p+\q)^2}{(\p-\q)^2}\;+\;h_F\;d_F\;\Bigr[\;
(-\epsilon_\p^2+{\hat q}_4^2\;)\;\ln(a_F^+a_F^-)
-d_F\;\ln\frac{a_F^+}{a_F^-}\;\Bigr]\;\right.\nonumber\\
&&\left.+\;\frac{1}{2}(\;h_F^2+d_F^2-4\p^2\q^2)\;\Bigr[\;
(-\epsilon_\p^2+{\hat q}_4^2\;)\;\ln\frac{a_F^+}{a_F^-}
-d_F\;\ln(a_F^+a_F^-)\;\Bigr]\;\right\}\nonumber \;,
\end{eqnarray}
\begin{eqnarray}
&&I_2=-e^2\int\limits_{0}^{\infty}
\frac{d|\p|}{4\pi^2}\;\frac{\p^2}{\epsilon_\p}\;
\frac{n_\p^++n_\p^-}{2}\frac{1}{4|\p||\q|(\epsilon_\p^2
+{\hat q}_4^2)^2}\left\{\;\Bigr[(-\epsilon_\p^2
+{\hat q}_4^2)(\p^2-\q^2)^2\;\right.\nonumber\\
&&\left.+(\epsilon_\p^2
+{\hat q}_4^2)^2(\p^2+\q^2) \Bigr]
\ln\frac{(\p+\q)^2}{(\p-\q)^2}\;+\;h_F\;d_F\;\Bigr[\;
(-\epsilon_\p^2+{\hat q}_4^2\;)\;\ln\frac{a_F^+}{a_F^-}
-d_F\;\ln(a_F^+a_F^-)\;\Bigr]\;\right.\nonumber\\
&&\left.+\;\frac{1}{2}(\;h_F^2+d_F^2-4\p^2\q^2)\;\Bigr[\;
(-\epsilon_\p^2+{\hat q}_4^2\;)\;\ln(a_F^+a_F^-)
-d_F\;\ln\frac{a_F^+}{a_F^-}\;\Bigr]\;\right\}\nonumber\;,
\end{eqnarray}
\begin{eqnarray}
&&I_3=-e^2\int\limits_{0}^{\infty}
\frac{d|\p|}{4\pi^2}\;\frac{n_\p^B}{|\p|}\;
\left\{\;h_B\;+\;\frac{1}{8|\p||q\|};\Bigr[\;
(h_B^2+d_B^2)\;\ln(a_B^+a_B^-)
+2d_B\;h_B\ln\frac{a_B^+}{a_B^-}\;\Bigr]\;\right\}\nonumber\;.
\end{eqnarray}
The abbreviations are the same as in Eq.(7). The integrals found are
essential for the next-to-leading orders beyond
$T^2$-approximation.  Here only the first integral in Eq.(55) is
useful to find $\mu/T$-corrections.

\newpage

\begin{center}
{\bf APPENDIX B}
\end{center}
Our starting point is the dispersion relation (3)
\begin {eqnarray}
[\;(iq_4-\mu)-{\bar K}_4]^2\;=\;\q^2\;(1+K)^2 +m^2(1+Z)^2
\end{eqnarray}
with $m\ne 0$ and we solve it in the $T^2$-approximation with $\mu\ne
0$. All calculations are performed in the Coulomb gauge taking into
account only the leading $T^2$-terms and $\mu/T$-corrections.
In this case the functions within Eq.(56) are:
\begin {eqnarray}
&&\!\!\!\!\!\!K(q_4,\q)\;=\;\frac{I_K}{\q^2}\Bigr(\;1+\frac{\xi}{2}\ln
\frac{\xi-1}{\xi+1}\Bigr)\;-\;I_B\;\Bigr(\;\xi-\frac{1}{2}
(1-\xi^2)\ln\frac{\xi-1}{\xi+1}\;\Bigr)\;\frac{1}{|\q|}\nonumber
\end{eqnarray}
\begin{eqnarray}
-{\bar K}_4(q_4,\q)\;=\;\frac{I_K}{2|\q|}
\ln\frac{\xi-1}{\xi+1}\;-\;I_B \,,\qquad
-Z(q_4,\q)\;=\;3I_Z\;+\;\frac{I_B}{|\q|}
\ln\frac{\xi-1}{\xi+1}
\end{eqnarray}
making it possible to essential simplify its solution. Here the variable
$\xi=\omega/|\q|$ is more convenient than $|\q|$ and as usual
$\omega=(iq_4-\mu)$. All integrals are defined by Eq.(12).

After the simple algebra has been performed within Eq.(56) our
result is the equation of the fourth degree with respect to
$\omega(\xi)$
\begin {eqnarray}
&&\omega^4\;[\;\xi^2-1\;]\;-\;2\omega^3\;\xi\;I_B\;[\;\xi-
b(\xi)\;]+\omega^2\;\xi^2\;[\;I_B^2\;(1-b(\xi)^2)
-m_R^2-2I_K\;]\nonumber\\
&&+\;2\omega\;\xi\;\;I_B\;[\;I_K\;(\;1+d(\xi)\;)\;b(\xi)\;+
\;\xi\;d(\xi)\;(\;2m_Rm-I_K\;)\;]\nonumber\\
&&-\;I_K^2\;\xi^4\;(\;1+d(\xi)\;)^2
\;+\;\xi^2\;d(\xi)^2\;(\;I_K^2-4m^2I_B^2\;)=0
\end{eqnarray}
where $m_R=m(1-3I_Z)$ is the renormalized fermion mass, and
the functions $d(\xi)$ and $b(\xi)$ are given by
\begin {eqnarray}
&&d(\xi)=\frac{\xi}{2}\;\ln\frac{\xi-1}{\xi+1}\nonumber\\
&&b(\xi)=\xi-\frac{1}{2}(1-\xi^2)\ln\frac{\xi-1}{\xi+1}\;.
\end{eqnarray}
The dispersion relation (58) being very complicated is not
solved analytically. However in the long wavelength limit (when
$\xi\rightarrow\infty$) it can be simplified as
\begin {eqnarray}
\Bigr[\omega^2-\omega(I_B-\eta m_R)-(I_K-2\eta mI_B)\Bigr]\cdot
\Bigr[\omega^2-\omega(I_B+\eta m_R)-(I_K+2\eta mI_B)\Bigr]=0
\end{eqnarray}
and one finds the rather simple solution
\begin {eqnarray}
\omega(0)=\frac{1}{2}\Big[\eta\;m_R+I_B\Big]\pm
\sqrt{\;\frac{[\eta\;m_R+I_B]^2}{4} +(I_K+2\eta mI_B)}
\end{eqnarray}
which demonstrates four well-separated effective masses:
two of them pertain to quasi-particle excitations and other
two to quasi-holes.  Here $\eta=\pm 1$, and the parameters
$m$ and $\mu$ are nonzero.

\newpage

\begin{center}
{\bf References}
\end{center}

\renewcommand{\labelenumi}{\arabic{enumi}.)}
\begin{enumerate}

\item{ A.~S.~Kronfeld, Phys.Rev. {\bf D58}, 051501 (1998).}

\item{ O.~K.~Kalashnikov and V.~V.~Klimov, Yad. Fiz. {\bf 31}, 1357
(1980) (Sov. J. Nucl. Phys. {\bf 31}, 699 (1980)).}

\item{ V.~V.~Klimov, Yad.Fiz. {\bf 33}, 1734 (1981) (Sov. J. Nucl.
Phys. {\bf 33}, 934 (1981)); Zh. Eksp. Teor. Fiz. {\bf 82}, 336 (1982)
(Sov.  Phys. JETP  {\bf 55}, 199 (1982)) .}

\item { H.~A.~Weldon, Phys. Rev. {\bf D26}, 2789 (1982).}

\item { O.~K.~Kalashnikov, Fortschr. Phys, {\bf 32}, 525 (1984).}

\item{ H.~A.~Weldon, Phys. Rev. {\bf D40}, 2410 (1989); Physica
{\bf A158}, 169 (1989).}

\item{ R.~D.~Pisarski, Nucl. Phys. {\bf A498}, 423c (1989).}

\item{ V.~V.~Lebedev and A.~V.~Smilga, Ann. Phys.(N.Y) {\bf
202}, 229 (1990).}

\item{ G.~Baym, J.~P.~Blaizot and B.~Svetitsky, Phys. Rev. {\bf D46},
4043 (1992).}

\item{ E.~Petitgirard,  Z. Phys.  {\bf C 54}, 673 (1992).}

\item{ C.~Quimbay and ~S.~Vargas-Castrillon, Nucl. Phys. {\bf B451},
265 (1995).}

\item{ E.~J.~Levinson and  D.~H.~Boal, Phys. Rev. {\bf D 31}, 3280
(1985).}

\item { J.~P.~Blaizot and J.~Y.~Ollitrault, Phys. Rev. {\bf D48},
1390 (1993).}

\item{ O.~K.~Kalashnikov, Mod. Phys. Lett. {\bf A12}, 347 (1997).}

\item{ O.~K.~Kalashnikov, Pis'ma  Zh. Eksp. Teor. Fiz. {\bf 67}, 3
(1998) (JETP Lett.{\bf 67}, 1 (1998)).}

\item{ O.~K.~Kalashnikov, Mod. Phys. Lett. {\bf A13}, 1719 (1998).}

\item {R.~Kobes,~G.~Kunstatter and A.~Rebhan, Nucl. Phys. {\bf B355},
1 (1991).}

\item {J.~C.~Breckenridge,~M.~J.~Lavelle,~T.~G.~Steele, Z.Phys.
{\bf C65}, 155 (1995) .}

\item{ O.~K.~Kalashnikov. Phys. Lett. {\bf B 279}, 367 (1992). }

\item{ E.~Braaten and R.~D.~Pisarski, Phys. Rev. {\bf D46}, 1829
(1992).}

\item{ I.~V.~Tyutin and V.~Zeitlin, Lebedev Inst. preprint
FIAN/TD/98-17, hep-th/9810058.}

\item{ O.~K.~Kalashnikov, Pis'ma  Zh. Eksp. Teor. Fiz. {\bf 41},
477(1985) (JETP Lett. {\bf 41}, 582 (1985)).}

\end{enumerate}

\end{document}